\documentclass{aa}
\usepackage{graphics}

\begin{document}

   \thesaurus{06         % A&A Section 6: Form. struct. and evolut. of stars
              (04.01.1;  %  Astronomical data bases: miscellaneous 
               04.01.2;  %  Atlases
               04.03.1;  %  Catalogs
               04.19.1)} %  Surveys.

   \title{The ALADIN Interactive Sky Atlas}

   \subtitle{A Reference Tool for Identification of 
           Astronomical Sources}

\author{Fran\c{c}ois Bonnarel
\and
   Pierre Fernique
\and
   Olivier Bienaym\'e
\and
    Daniel Egret
\and
   Fran\c{c}oise Genova
\and
   Mireille Louys\thanks{Laboratoire des Sciences de l'Informatique,
    de l'Image et de la T\'el\'ed\'etection, ENSPS, Universit\'e
   Louis Pasteur, F-67000  Strasbourg, France}
\and
   Fran\c{c}ois Ochsenbein
\and 
   Marc Wenger
\and
   James G. Bartlett\thanks{\emph{Present address:} Observatoire
       Midi-Pyr\'en\'ees, Toulouse, France}
%\and
%    Michel Cr\'ez\'e\thanks{\emph{Present address:} IUP de Vannes,
%       Tohannic, rue Yves Mainguy, F-56000 Vannes, France}
          }

   \offprints{Daniel Egret}
   \mail{question@simbad.u-strasbg.fr}

   \institute{CDS, Observatoire astronomique de Strasbourg, UMR 7550,
11 rue de l'Universit\'e, F-67000 Strasbourg, France}

   \date{Received 6 December 1999 / Accepted 16 December 1999}

\maketitle

   \begin{abstract}

  The {\sc Aladin} interactive sky atlas, developed at CDS, 
is a service providing simultaneous access to digitized
images of the sky, astronomical catalogues, and databases.
   The driving motivation is to facilitate direct, 
visual comparison of observational data at any wavelength with 
images of the optical sky, and with reference catalogues.  

   The set of available sky images consists of the STScI Digitized
Sky Surveys, completed with high resolution images of crowded
regions scanned at the MAMA facility in Paris.

    A Java WWW interface to the system is available at:

{http://aladin.u-strasbg.fr/}.


\end{abstract}

%
%________________________________________________________________

\section{Introduction}

\subsection{The CDS}
The Centre de Donn\'ees astronomiques de Strasbourg
(CDS) defines, develops, and maintains services
to help astronomers find the information
they need from the
very rapidly increasing wealth of astronomical information, 
particularly on-line information. 

In modern astronomy,  cross-matching data acquired at
different wavelengths is often the key  to the understanding of
astronomical phenomena, which means that astronomers have to use data
and information produced in fields in which they are not specialists.
The development of tools for cross-identification of objects 
is of particular importance in this context of 
multi-wavelength astronomy. 

A detailed description of the CDS on-line services can be found, e.g., 
in Egret et al.\ (\cite{cds-amp2})
and in Genova et al.\ (\cite{cds-hub}, \cite{cds}, \cite{cds2000}), 
or at the CDS web site\footnote{\emph{Internet address:}
http://cdsweb.u-strasbg.fr/}.

\subsection{The ALADIN Project}

%%%%
Several sites currently provide on-line access 
to digitized sky surveys at different wavelengths:
this is, for instance, the case of
Digitized Sky Survey (DSS) at STScI 
(Morrison \cite{1995adass...4..179M}),
and of similar implementations at other sites,
providing quick access to cutouts of the 
compressed DSS images.
{\sc SkyView} at HEASARC (McGlynn et al. \cite{skyview}) 
can generate images of any portion 
of the sky at wavelengths in all regimes from
radio to gamma-ray.
%%%%%
Some of these services  provide
simultaneous access to images
and to catalogue data.
The {\sc SkyCat} tool, recently developed
at ESO (Albrecht et al.\ \cite{skycat}),
addressed this concern in the context of 
the European Southern Observatory scientific environment 
(in view of supporting future users of the Very Large Telescope);
{\sc SkyCat} uses a standardized syntax to access heterogeneous
astronomical data sources on the network.

{\sc Aladin} has been developed independently by the CDS
since 1993
as a dedicated tool for identification
of astronomical sources -- a tool that can fully benefit from the
whole environment of CDS databases and services, and that is designed
in view of being released as a multi-purpose
service to the general astronomical community.

{\sc Aladin} is an interactive sky atlas, 
allowing the user to visualize a part of the sky,
extracted from a database of images
from digitized surveys and observational
archives, 
and to overlay objects
from the CDS catalogues and tables, 
and from reference databases ({\sc Simbad} and NED),
upon the digitized image of the sky. 

It is intended to become a major cross-identification
tool, since it allows recognition of astronomical
sources on the images at optical
wavelength, and at other wavelengths through the catalogue data. 
Expected usage scenarios include multi-spectral 
approaches such as searching for counterparts of 
sources detected at various  wavelengths, and 
applications related to 
careful identification of astronomical objects.
%%%%%
{\sc Aladin} is also heavily used for the
CDS needs of catalogue and database quality control. 

In the case of extensive undertakings (such as 
checking the astrometric
quality for a whole catalogue), it is expected
that {\sc Aladin} will be useful for understanding the
characteristics of the catalogue or survey, and for setting
up the parameters to be adjusted while fine tuning the
cross-matching or classification algorithms, by studying
a sample section of objects or fields.

A discussion of the usage of such a tool for cross-identification 
can be found in Bartlett \& Egret (\cite{xid-179}),
where it is shown how \emph{training sets} are used to
build likelihood ratio  tests. 

The {\sc Aladin} interactive atlas is available in three modes:
a simple previewer, a Java interface, and an
X-Window interface.  We describe here mostly the Java interface
which is publicly accessible on the World-Wide Web.

%
%______________________________________________________________

\section{Access modes}

After a long phase of development, 
(see e.g., Paillou et al.\ \cite{paillou}),
{\sc Aladin} has been first distributed to a
limited number of astronomy laboratories in 1997,
as an X-Window client program, 
to be installed on a Unix machine on the user side.
The client program interacts with the servers running
on Unix workstations at CDS (image server, catalogue
server, {\sc Simbad} server) and manages image handling
and plane overlays.

The strategy of having a client program on the user side
is difficult to maintain on the long run.
The World-Wide Web offers, with the development of Java
applications (or \emph{applets}), a way to solve this difficulty.  
Actually, there is still a \emph{client} program:
this is the Java applet itself, 
that the user receives from the WWW server.
Most current Internet browsers are able to make it run
properly, so that the user does not have to install anything
special other than an Internet browser.

As a consequence, {\sc Aladin} is currently available 
in the three following modes:

\begin{description}
\item[{\sc Aladin} previewer:] a pre-formatted
image server provides a compressed image of fixed size
($14.1\arcmin \times 14.1\arcmin$ for the DSS-I)
around a given object or position.  When an
object name is given, its position is resolved
through the {\sc Simbad} name resolver.  Anchors
pointing to the previewer are integral part of 
the World-Wide Web interfaces
to the {\sc Simbad} database\footnote{\emph{Internet
address}: http://simbad.u-strasbg.fr/Simbad}
and to the CDS bibliographic service\footnote{\emph{Internet
address}: http://simbad.u-strasbg.fr/biblio.html}.
%%%%
The result page also gives access to the full resolution
FITS image for download.

\item[{\sc Aladin} Java:]  this is the primary
public interface, supporting queries to the
image database and overlays from any catalogue or table
available at CDS, as well as from {\sc Simbad} and NED databases.
Access to personal files is not possible (due to
security restrictions of the Java language).
These restrictions do not apply to the \emph{stand-alone}
version, which can be installed and run on a local 
\emph{Java virtual machine}. 

\item[{\sc Aladin~X}:] The X-Window {\sc Aladin} client
provides most of the functionalities of the {\sc Aladin} Java
interface, plus more advanced functions, as described
below (section~\ref{AladinX}).

\end{description}

%
%______________________________________________________________

\section{The image database}

\subsection{Database summary}

The {\sc Aladin} image dataset consists of:

\begin{itemize}

\item  The whole sky image database from the first Digitized
Sky Survey (DSS-I) digitized from photographic plates
and distributed by the Space Telescope Science
Institute (STScI) as a set of slightly compressed
FITS images  (with a resolution of $1.8\arcsec$);
DSS-II is also currently
being integrated into the database (see below);

\item  Images of \emph{crowded} fields
(Galactic Plane, Magellanic Clouds) at the full resolution 
of $0.67\arcsec$, scanned at the {\em Centre d'Analyse des
Images} (MAMA machine) in Paris; 

\item  Global plate views 
($5\degr \times 5\degr$
or $6\degr \times 6\degr$ according to the
survey) are also available for all the plates
contributing to the image dataset:
these are built at CDS by averaging
blocks of pixels from the original scans;

\item  Other images sets, or user-provided images,
in FITS format, having suitable World Coordinate System 
information in the header (see e.g.\ Greisen \& Calabretta \cite{wcs});
this functionality is currently available only for the
Java stand-alone version.
\end{itemize}

\subsection{Building the database contents}

The {\sc Aladin} project has set up collaborations with 
the major groups providing digitizations of sky surveys.
The original surveys are made of photographic Schmidt plates
obtained at Palomar in the North, 
and ESO/SERC in the South, and
covering the whole sky at different epochs and colours
(see e.g., MacGillivray \cite{potsdam}).  

The database currently includes the first Digitized Sky 
Survey (DSS-I) produced by the
Space Telescope Institute (Lasker \cite{dss}), 
for the needs of the Hubble
Space Telescope.
To create these images, the
STScI team scanned the first epoch (1950/1955) Palomar 
$E$ Red and
United Kingdom Schmidt $J$ Blue plates 
(including the SERC J Equatorial Extension and some
short V-band plates at low galactic latitude) 
with a pixel size of $1.7\arcsec$ ($25{\mu}m$).
The low resolution and a light data compression
(factor of 10) permit storage of images covering
the full sky on a set of 102 CD-ROMs.

%%%%%%
DSS-II images in the R-band (from Palomar POSS-II F and
UK Schmidt SES, AAO-R, and SERC-ER), scanned with
a $1\arcsec$ ($15{\mu}m$) sampling interval 
(see Lasker \cite{dss-ii}) 
are gradually being included into the system,
and will soon be followed by DSS-II images
in the B-band (POSS-II J). 

In addition, high resolution digitalization
of POSS-II, SERC-J, SERC-SR, SERC-I, or ESO-R plates 
featuring crowded regions of the sky (Galactic Plane
and Magellanic Clouds) have been provided by the MAMA
facility at the {\sl Centre d'Analyse des Images} (CAI),
Observatoire de Paris (Guibert \cite{MAMA}). 
Sampling is $0.67\arcsec$ per pixel ($10{\mu}m$).
Currently, these high resolution images cover
about 15\% of the sky, and are stored in a juke-box
of optical disks, with a capacity of 500 Gigabytes.

\subsection{The image server}

The image server for {\sc Aladin} had to be able
to  deal with various survey data,
in heterogeneous formats (uncompressed FITS, compressed
JPEG or PMT -- see Section \ref{compression}, etc.).
For that, an object-oriented design was chosen,
allowing an easy manipulation of image calibrations
and headers, through the use of object classes.
Image compression or decompression, image reconstruction,
and in a near future, part of the recalibration, are 
seen as class methods.

Images are currently divided into subimages of
$500 \times 500$ pixels (DSS-I),
$768 \times 768$ pixels (DSS-II),
or $1024 \times 1024$ pixels (MAMA).

The 1.5 million subimages are described by
records stored in a relational database,
encapsulated by several classes of the image
management software.
When an image of the sky is requested,
the original subimages containing the
corresponding sky area are retrieved
through SQL commands, and 
the resulting image is built on the fly.

%
%______________________________________________________________
\section{Usage scenarios}

In this section we will focus on describing the usage
of the {\sc Aladin} Java interface, as it is available now 
(November 1999).

\subsection{Access}

The {\sc Aladin} home page is available through
the CDS Web server at the following address: 
http://aladin.u-strasbg.fr/

This site provides access to {\sc Aladin} documentation,
including scientific reports, recent publications, etc.

\subsection{Query modes}

The typical usage scenario starts with a request
of the digitized image for an area
of the sky defined by its central position or name of 
central object (to be resolved by {\sc Simbad}).
The size of the sky field is determined by the
photographic survey used: it is $14.1\arcmin$ in
the case of the DSS-I.

Astrometric information comes from 
the FITS header of the DSS image, and is generally
accurate to the arcsecond (with deviations up
to several arcsec.\ in exceptional cases, on plate edges).

\begin{figure}
%\vspace{\hsize}
\resizebox{\hsize}{!}{\includegraphics{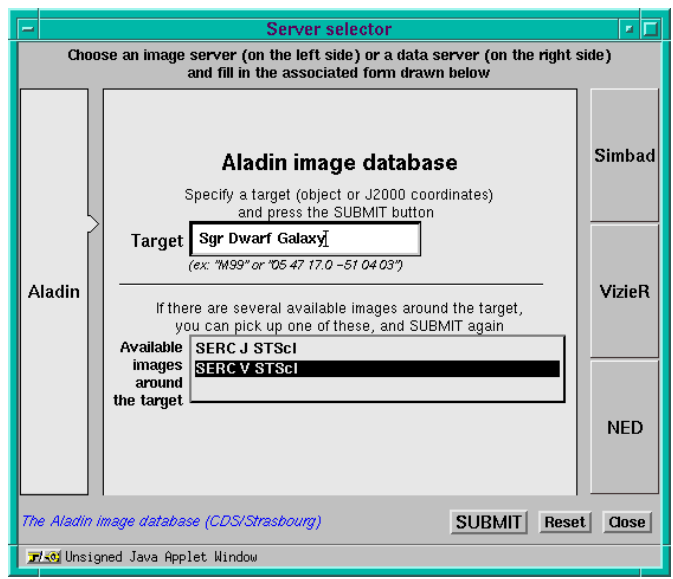}}
\caption{Example of Images/Data {\sc Aladin} query panel.}
\label{query-panel}
\end{figure}

\begin{figure*}
%\vspace{\hsize}
% 18cm if double column
\resizebox{\hsize}{!}{\includegraphics{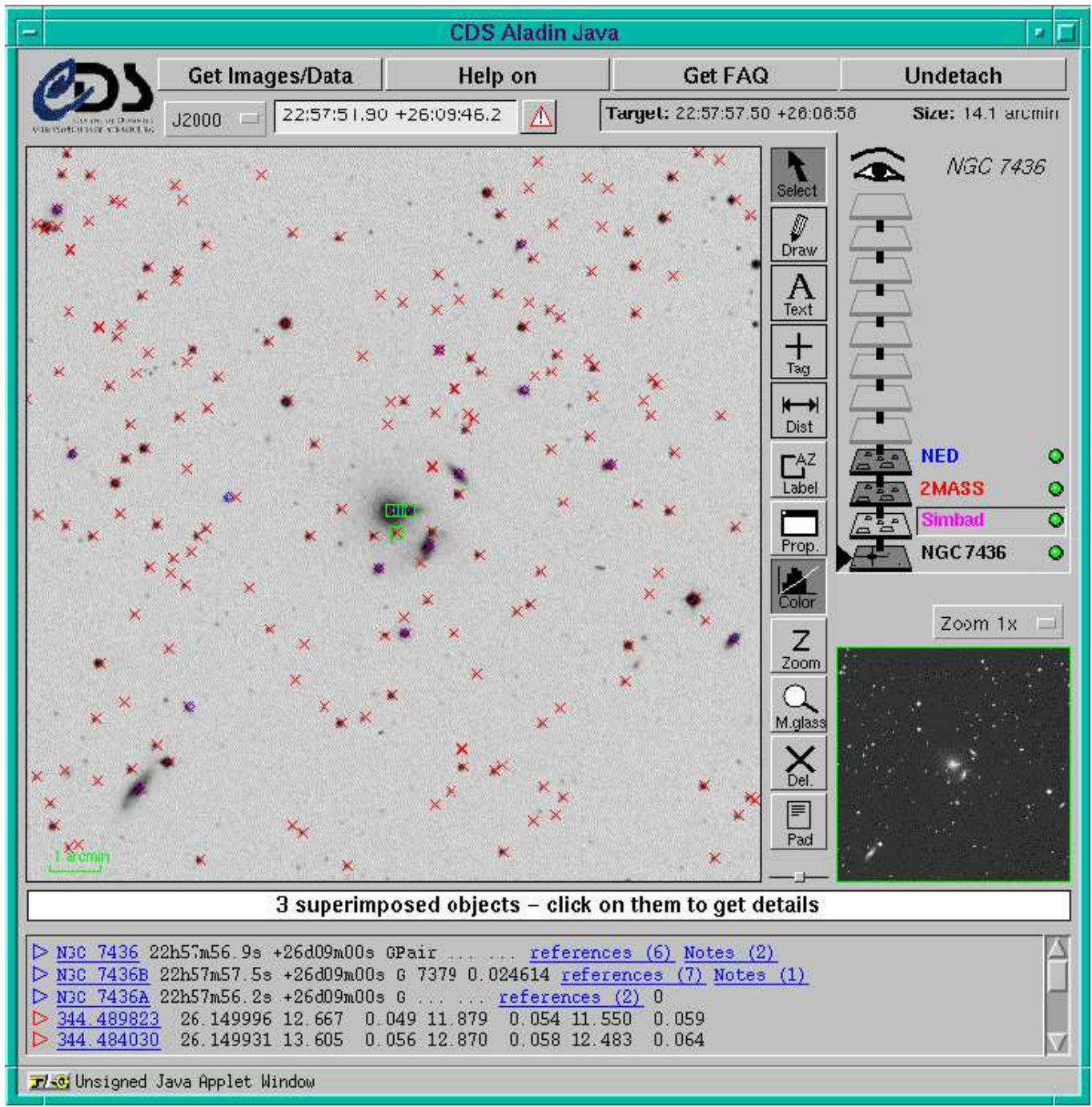}}
\caption{Example of {\sc Aladin} window, with an image centered
on NGC 7436, and objects from NED and 2MASS marked by symbols.}
\label{sample-image}
\end{figure*}

In a subsequent step, the interface, 
illustrated by Figs.~\ref{query-panel} and \ref{sample-image},
allows the user to stack several information
planes related to the same sky field, to superimpose the
corresponding data from catalogues and databases, 
and to obtain interactive access to
the original data.

The possible information planes are the following:

\begin{itemize}

\item  Image pixels from the {\sc Aladin} database of
digitized photographic plates (DSS-I, MAMA, DSS-II);
functionalities include zooming capabilities, inverse
video, modification of the color table;

\item  Information from the {\sc Simbad} database
(Wenger et al.\ \cite{simbad}); objects
referenced in {\sc Simbad} are
visualized by color symbols overlaid on top of the black and
white image; the shape and color of the symbols can be
modified on request, and written labels can be added
for explicit identification of the objects; these 
features are also available for all
the other information planes; 

\item  Records from the CDS library of catalogues
       or tables ({\sc VizieR}\footnote{\emph{Internet address:}
      http://vizier.u-strasbg.fr/}, Ochsenbein et al.\
     \cite{vizier});
    the user can select the desired catalogue from a
preselected list including the major reference catalogues
such as the Tycho Reference Catalogue 
(ESA \cite{tyc}; H{\o}g et al.\ \cite{trc}), 
GSC (Lasker et al.\ \cite{gsc}), IRAS Point Source Catalog,
or USNO A2.0 (Monet \cite{usno}); the user can alternatively
select the catalogues for which entries may be available
in the corresponding sky field, using the {\sc VizieR} 
query mechanism by position
(see \ref{catserver}), catalogue name or keyword; 

\item  Information from the NED database: objects
referenced in the NASA/IPAC Extragalactic 
Database\footnote{\emph{Internet address:} 
http://nedwww.ipac.caltech.edu/}
(Helou et al. \cite{ned})
can also be visualized through queries submitted to the
NED server at IPAC;

%%%%%%
\item   Archive images will gradually become available
through the corresponding mission logs: Hubble Space Telescope
images are currently available 
(see Fig.~\ref{M16} for an example),
and more archives will follow.

\begin{figure*}
%\vspace{\hsize}
% 18cm if double column
\resizebox{\hsize}{!}{\rotatebox{+90}{\includegraphics{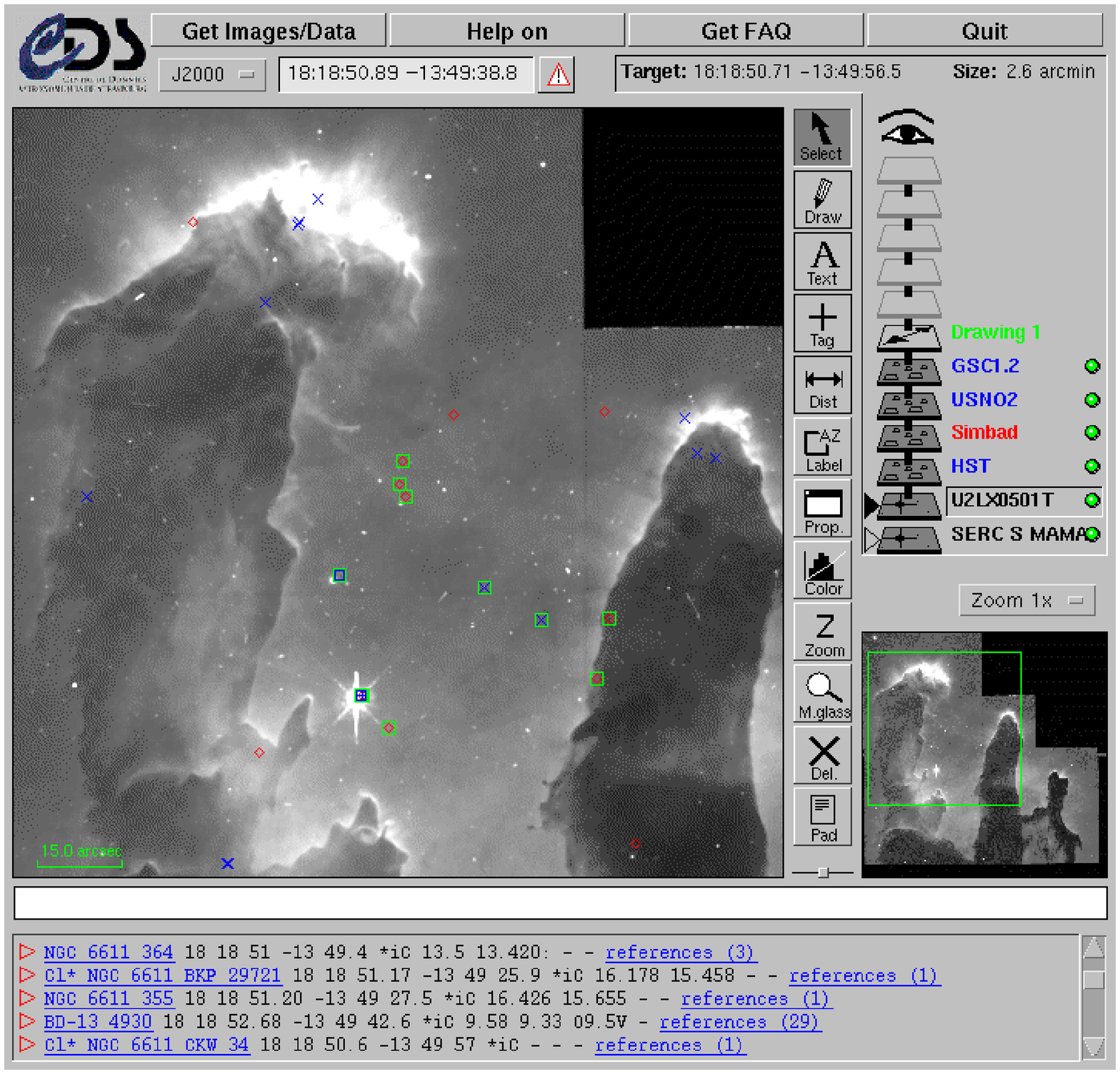}}}
\caption{Example of {\sc Aladin} display of a famous image
from the HST archive (WFPC2) featuring the Eagle Nebula
(\object{M 16}).
Objects present in Simbad, GSC, and USNO A2.0 are flagged
with different symbols. Field size is 2.6 arcmin
(full image, right) and 1.6 arcmin (left).}
\label{M16}
\end{figure*}

\item  Local, user data files can also be overlaid,
but, because of current limitations of the Java applications,
this feature is only available in the stand-alone
version, or in {\sc Aladin~X}.
\end{itemize}
 
The stack of images and graphics is made visible
to the user (under the eye icon, on the
right of Fig.~2) so that each plane can be 
turned on and off. The status of queries
is also easily visualized.

For all information planes ({\sc Simbad}, {\sc VizieR}, NED) links are
provided to the original data. This is done in the following
way: when selecting an object on the image,
with mouse and cursor, it is
possible to call for the corresponding information which
will appear in a separate window on the Internet browser.
It is also possible to select with the mouse and cursor all
objects in a rectangular area: the corresponding
objects are listed in a panel on the bottom of the {\sc Aladin}
window;  this list includes basic information (name,
position and, when applicable, number of bibliographical
references) and anchors pointing to the original catalogue 
or database.

At any moment the position of the cursor is translated
in terms of right ascension and declination on the sky
and visualized in the top panel of the {\sc Aladin} window.
Additional features are available, such as a tool for
computing angular distance between marked objects.

%%%%%
The \emph{standalone} version gives
access to additional facilities, 
not available through the Web, such as printing and
saving the images and data.

\subsection{The catalogue server}
\label{catserver}

The ability to access all {\sc VizieR} catalogues and tables 
directly from {\sc Aladin} is a unique feature
which makes it an extremely powerful tool for 
any cross-identification or classification work.

The ``\emph{Select around target}'' request relies on
a special feature -- the genie of the lamp:
this is the ability to decide which catalogues, among
the database of (currently) over 2,600 catalogues
or tables, contain data records for astronomical objects
lying in the selected sky area.
In order to do that, an index map of {\sc VizieR} catalogues
is produced (and kept up-to-date), on the basis of about 
ten pixels per square degree: for each such `pixel' the
index gives the list of all catalogues and tables
which have entries in the field.

When a user hits the button ``\emph{Select around target}'',
this index is queried and the list of useful
catalogues is returned. It is possible, at this stage,
either to list all catalogues, or to produce a subset
selected on the basis of keywords.
Note that, as the index ``pixels'' generally match
an area larger than the current sky field,
there is simply a good chance, but not 100\%, 
to actually obtain
entries in the field when querying one of the selected tables.

\subsection{Cache}

The images of the 30,000 most cited objects in {\sc Simbad} 
are pre-computed
and available on a cache on magnetic disk. 
For these objects, the image is served much faster
than for other objects
where the image has to be extracted
from the Digitized Sky Survey.

%%%%%%
\subsection{Usage statistics}

As the newest service developed by CDS, {\sc Aladin} has
not yet been widely publicized, and its usage is in
a steeply growing phase. Currently about
10,000 queries  are processed monthly,
generating the extraction of more than 5,000 images.

%______________________________________________________________
\section{Image compression}
\label{compression}

Astronomical image compression in
the context of {\sc Aladin} has been discussed 
in detail by  Louys et al.\ (\cite{louys}).

For the {\sc Aladin} Java interface and for the {\sc Aladin}
previewer, the current choice has been to deliver to the
user an image in  JPEG 8-bit format, constructed from the original
FITS images. JPEG is a general purpose standard
which is supported by all current Internet browsers.
The size of such an image does not exceed 30 kBytes, and
thus the corresponding network load is very small.

In the near future,
the Pyramidal Median Transform (PMT) algorithm,
implemented in the MR-1 package
(Starck et al.\ \cite{pmt}), 
will be used within {\sc Aladin}
for storing or transferring new image datasets,
such as additional high resolution images 
(see again Louys et al.\ \cite{louys} for details).
%%%%%
The corresponding decompression package is
being written in Java code, and could be downloaded
on request for use within the Java interface.

%______________________________________________________________
\section{Aladin~X}
\label{AladinX}

The {\sc Aladin} X-Window interface is 
the testbed for further developments.
It is currently only distributed for the Unix Solaris
operating system.
Interested potential users should contact CDS for details.

\subsection {Source extraction}

{\sc Aladin~X} includes a procedure for source extraction.
The current mechanism will soon be
replaced by SExtractor (Bertin \& Arnouts
\cite{1996A&AS..117..393B}).

\subsection{Plate calibrations}

While the first level astrometric calibrations are given
by the digitizing machines, a second level 
is being developed that
will allow the user to \emph{recalibrate} the image with
a new set of standards taken, for example, from the
Tycho Reference Catalogue.
The photometric calibrations (surface and
stellar) will eventually also be performed within {\sc Aladin}, by using
the Guide Star Photometric Catalogs (GSPC I and II;
Ferrari et al.\ \cite{gspc2}; Lasker et al.\ \cite{gspc1}).

Users will thus be able to work on the details of local astrometric
and photometric plate calibrations in order to
extract the full information from the digitized plates.

%______________________________________________________________
\section{Integration of distributed services}

While the CDS databases have followed different 
development paths, the
need to build a transparent access 
to the \emph{whole set} of CDS services has become
more and more obvious with the easy
navigation permitted by hypertext tools. 
{\sc Aladin} has become the prototype of such a development,
by giving comprehensive simultaneous 
access to {\sc Simbad}, 
the {\sc VizieR} Catalogue service, 
and to external databases such as NED,
using a client/server approach and, when possible,
standardized query syntax and formats. 

In order to be able to go further, the
CDS has built a general data
exchange model, taking into account all types of information available
at the Data Center, known under the acronym
of GLU for G\'en\'erateur de Liens Uniformes -- Uniform
Link Generator (Fernique et al.\ \cite{glu}). 

More generally, with the development of the Internet, 
and with an increasing
number of on-line astronomical services giving access 
to data or information, it has become critical to develop
new tools providing access to distributed services. This
is, for instance, the concern expressed by NASA through 
the AstroBrowse project (Heikkila et al.\ \cite{astrobrowse}).
A local implementation of this concept is available at CDS 
(AstroGlu:  Egret et al.\ \cite{astroglu}).

%
%______________________________________________________________

\section{Future developments}

An important direction of development in the near future
is the
possibility of providing access to images from other sky surveys
or deep field observations: obvious candidates are
the DENIS (Epchtein \cite{denis}) and 2MASS
(Skrutskie \cite{2mass}) near-infrared surveys.
%%%%%
The first public point source catalogues 
resulting from these surveys
are already available through {\sc Aladin},
since they are included in the {\sc VizieR}
service. This has already proved useful for validating
survey data in  preliminary versions of the DENIS
catalogue (Epchtein et al. \cite{denis-psc}).

The CDS team will also continue to enrich 
the system functionality.
The users play an important role in
that respect, by giving feedback on the desired features
and user-friendliness of the interfaces.

%%%%
New developments are currently considered
as additional modules which will be incorporated to the
general release only when needed, possibly
as optional downloads, in order to keep
the default version simple and efficient enough for 
most of the Web applications.

On a longer term, the CDS is studying the possibility of
designing \emph{data mining} tools that will help to make a
fruitful use of forthcoming very large surveys, and will
be used for cross-matching several surveys obtained, 
for instance, at different wavelengths.
A first prototype, resulting from a collaboration between
ESO and CDS, in the framework of the VLT scientific
environment is currently being implemented (Ortiz et al.
\cite{ortiz}).

%
%______________________________________________________________

%\{Conclusion}
%
%The {\sc Aladin} tool  is designed to be an essential tool for
%multi-wavelength cross-identifications.

\begin{acknowledgements}
CDS acknowledges the support of INSU-CNRS, the Centre National
d'Etudes Spatiales (CNES), and Universit\'e Louis Pasteur.

We are indebted to Michel Cr\'ez\'e who initiated
the project while being Director of the CDS, and to 
all the early contributors to the {\sc Aladin} project:
Philippe Paillou, 
Joseph Florsch, 
Houri Ziaeepour, 
Eric Divetain,
Vincent Raclot. 

Collaboration with STScI, and especially with the
late Barry Lasker, and with
Brian McLean, is gratefully acknowledged.
We thank Jean Guibert and Ren\'e Chesnel from CAI/MAMA
for their continuous support to the project.

The Digitized Sky Survey was produced at the Space Telescope Science 
Institute under U.S. Government grant NAG W-2166. 
The images of these surveys are based on photographic data obtained 
using the Oschin Schmidt
Telescope on Palomar Mountain and the UK Schmidt Telescope.

Java is a registered trademark of Sun Microsystems.

\end{acknowledgements}

%\listofobjects

\end{document}